\documentclass[showpacs,twocolumn,pra,superscriptaddress]{revtex4}
\usepackage{amsmath}
\usepackage{graphicx}
\usepackage{dcolumn}
\usepackage{color}
\usepackage{epsfig}
\usepackage{epstopdf}
\usepackage{bm}

\newcommand{\ket}[1]{\left\vert#1\right\rangle}
\newcommand{\bra}[1]{\left\langle#1\right\vert}
\graphicspath{./figs/}

\begin{document}

\title{Routing quantum information in spin chains}

\author{Simone Paganelli} \email{pascualox@gmail.com}
\affiliation{International Institute of Physics, Universidade Federal do Rio Grande do Norte, 59012-970 Natal, Brazil}
\affiliation{Departament de F\'{i}sica, Grup de F\'{i}sica Te\`{o}rica: Informaci\'{o} i Fen\`{o}mens Qu\`{a}ntics, Universitat Aut\`{o}noma de Barcelona, E-08193 Bellaterra, Spain}

\author{Salvatore Lorenzo}
\affiliation{Dipartimento di  Fisica, Universit\`a della Calabria,
87036 Arcavacata di Rende (CS), Italy} \affiliation{INFN - Gruppo
collegato di Cosenza}

\author{Tony~J.~G.~Apollaro}
\affiliation{Dipartimento di  Fisica, Universit\`a della Calabria,
87036 Arcavacata di Rende (CS), Italy} \affiliation{INFN - Gruppo
collegato di Cosenza} \affiliation{Centre for Theoretical Atomic,
Molecular and Optical Physics, School of Mathematics and Physics,
Queen's University, Belfast BT7 1NN, United Kingdom}

\author{Francesco Plastina}
\affiliation{Dipartimento di  Fisica, Universit\`a della Calabria,
87036 Arcavacata di Rende (CS), Italy} \affiliation{INFN - Gruppo
collegato di Cosenza}

\author{Gian Luca Giorgi} 
\affiliation{Department of Physics, University College Cork, Cork, Republic of Ireland}
\affiliation{AG Theoretische Quantenphysik, Theoretische Physik, Universit\"at des Saarlandes, D-66123 Saarbr\"ucken, Germany}

\pacs{03.67.Hk, 03.67.Pp, 75.10.Pq}

\begin{abstract}
Two different models are presented that allow for efficiently performing routing of a quantum state. Both cases
involve an $XX$ spin
chain working asspin chain working as a data bus and additional spins that play the role of sender and receivers,
one of which is selected to be the target of the quantum state transmission protocol via a coherent quantum
coupling mechanism making use of local and/or global magnetic fields. Quantum routing is achieved in the first
of the models considered by weakly coupling the sender and the receiver to the data bus. On the other hand, in
the second model, local magnetic fields acting on additional spins located between the sender and receiver and
the data bus allow us to perform high-fidelity routing.\end{abstract}
\maketitle

\section{Introduction}\label{intro}

The development of quantum technologies relies on the
ability to establish correlations between distant parties~\cite{nielsen}.
Whereas, photons are ideal carriers of quantum information
in free space~\cite{cirac97} since they interact weakly with the external
environment, solid-state devices are probably more suitable
for quantum communication within a computer.

In particular, spin chains with nearest neighbor interaction offer
a wide range of solutions for Quantum State Transfer (QST)
protocols~\cite{bose2003,rev}. Apart from their simple theoretical
description, they can be efficiently implemented in arrays of
trapped ions~\cite{porras04,deng,johanning} or by using cold atoms in 
optical lattices~\cite{duan,greiner,sherson}, where single spin
addressing has been recently reported  \cite{weitenberg}.

Since, in QST protocols, the initial state is usually confined to a
small region of space, its transmission through a long unmodulated 
chain will inevitably  involve all of the modes of the
chain itself. As a consequence, state reconstruction in a
different spatial location will be affected by the detrimental
dispersion the spin wave packet is subjected to. Various proposals
have been made to overcome this drawback. In Ref.~\cite{datta},
the authors suggested using engineered spin-spin coupling and
found a way to obtain perfect QST independent of the chain
length. However, such an implementation would require a high
degree of control of the internal structure of the system, which
is not desirable from the experimental point of view. Alternative
methods are based on the use of trapped topological
fields~\cite{fdp}, on the extension of the encoding to more than
one site~\cite{linden}, on the use of strong dynamically switched
on interactions between the sender and the receiver with the
bus~\cite{banchi2011}.

One of the more explored solutions consists of weakly
coupling the sender and the receiver to the bulk chain. 
Roughly speaking, the resultingQST takes place in two distinct regimes:
For very weak coupling, the bulk chain behaves merely like an
information bus without being appreciably populated, and the
probability amplitude of finding the excitation undergoes an
effective Rabi oscillation between the sender and the 
receiver~\cite{wojcik,bus,wojcik1,campos,yao}; whereas, for nonperturbative end-point couplings,
the relevant modes taking part in the quantum state dynamics
reside mainly in the linear zone of the spectrum, thus,
minimizing the effect of dispersion so that QST occurs in
the so-called ballistic regime~\cite{BACVV11,tony, BanchiV13}.

A step beyond QST is represented by the possibility of routing
information from one sender to many possible receivers
with minimal control of the system; that is, without modifying
any of the spin-spin coupling parameters of the Hamiltonian.
Achieving this goal would clearly increase the degree of
connectivity of a spin bus by allowing the possibility to couple
the quantum node of a spin network to many receivers.

Despite the large number of papers on QST involving one
sender and one receiver, there are relatively few papers on
quantum routing. Actually, a setup admitting QST from a
sender to a single receiver may not be trivially extended to
implement a routing scheme: By way of example,
in Ref.~\cite{Kay11} it is explicitly demonstrated that perfect quantum state routing
is forbidden unless experimentally demanding operations
or severe Hamiltonian engineering is performed. Even by
relaxing the request of perfect QST, the problem still remains
nontrivial, especially in the huge class of QST protocols based
on mirror symmetry where a pivotal role is played by matrices
being both persymmetric and centrosymmetric~\cite{BanchiV13}.

It is the aim of this paper to discuss the dynamical behavior
of two coupling schemes that explicitly allow for an efficient
routing to be performed.

Previous proposal in this direction were formulated by Zueco {\it
et al.} in Ref.~\cite{Zueco2009} and Bose {\it et al.} in
Ref.~\cite{BoseJK2005}. In the former reference, the authors
considered an $XY$ chain in the presence of an external magnetic field harmonically
oscillating in time and two possible receivers; whereas, in the
latter, by exploiting the Aharonov-Bohm effect, high-fidelity
three-party communication has been shown to be achievable.
Routing between distant nodes in quantum networks has been
proposed in Refs.~\cite{Nikolopoulos08,BroughamNJ09}
where perfect QST is investigated
in a dual-channel quantum directional coupler and in a passive
quantum network, respectively, and in Refs.~\cite{kay,chudzicki} 
iin the presence of local control of the network nodes. A scheme
for routing entanglement has been also proposed in coupled
two-impurity channel Kondo systems\cite{BayatBS2010}.

Here, instead, we propose two different quantum router
protocols, which can be performed in $XX$ spin chains and in
which the local energies of the receiver do not need any control
or manipulation during the whole process.

The paper is organized as follows. In Sec.~\ref{sec:model}  we introduce
the first routing scheme where the sender and receivers interact
weakly with a spin ring and efficient QST is enabled by
coupling resonantly the sender with a chosen receiver bymeans
of a suitably chosen magnetic field. In Sec.~\ref{sec:2model}, the sender and
receivers are not directly coupled to the spin bus but rather
via effective Òbarrier qubits,Ó on which strong magnetic fields
act as knobs for the QST. In this latter scheme, a uniformly
coupled spin chain is considered, thus, avoiding the need for
bond control. Finally, in Sec.~\ref{sec:concl}, conclusions
are drawn, and future perspectives are discussed.

\section{Quantum router via weak bonds}\label{sec:model}
\begin{figure}
\begin{center}
\includegraphics[width=8cm,angle=0]{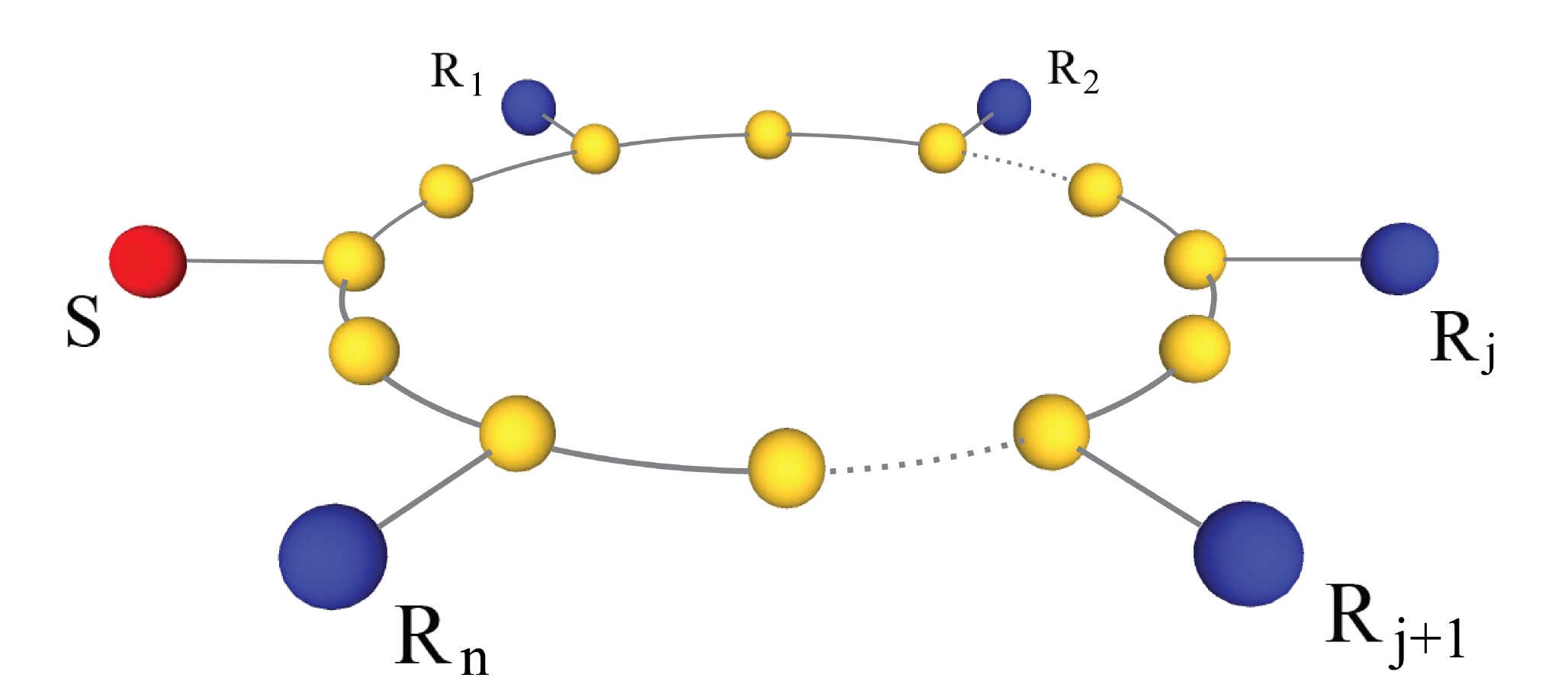}
\caption{(Color online) Schematic of the quantum
router. The sender and the receivers are coupled to a common chain
that acts as a quantum data bus.}\label{fring}
\end{center}
\end{figure}

Let us consider $N$ spins embedded in an $XX$ chain in the presence
of a transverse field plus $n+1$ spins (one sender and $n$
receivers) locally connected to the chain. A pictorial view of
this model is given in Fig.~\ref{fring}. The total Hamiltonian,
describing the chain, the sender and receivers, and their
coupling, respectively, reads $H{=}H_{C}+H_{I}+H_{CI}$, where
\begin{eqnarray*}
   H_{C} &{=}&-J\sum_{l{=}1}^N (\sigma^x_l \sigma^x_{l+1}+\sigma^y_l \sigma^y_{l+1})-h\sum_{l{=}1}^N \sigma^z_l,\\
    H_{I}&{=}& - h_S\sigma^z_S-  \sum_{i=1}^n h_{ R_{i}}\sigma^z_{R_{i}},\\
H_{CI}&{=}& - \frac{g}{2} (\sigma^x_{l_S} \sigma^x_{S}+\sigma^y_{l_S} \sigma^y_{S})- \frac{g}{2}
\sum_{i=1}^n (\sigma^x_{l_{R_{i}}} \sigma^x_{R_{i}}+\sigma^y_{l_{R_{i}}} \sigma^y_{R_{i}}).
\end{eqnarray*}
We have labeled the chain sites with $l{=}1,2, \dots,N$,  whereas
$S$ stands for the sender and $\vec R\equiv \{R_1,
R_2,\dots,R_n\}$ identifies the location of  the $n$ receivers.
The chain site $l_S$ is coupled to the sender, whereas the  $i$th
receiver is coupled to the site $l_{ R_{i}}$. Boundary conditions
are imposed by assuming $\sigma^\alpha_{N+1}{=}\sigma^\alpha_{1}$
($\alpha=x,y,z$). In the single excitation subspace, which will be
used henceforth, an exact mapping can be performed by relating
spin operators to fermion annihilation and creation operators. The
mapping consists of $\sigma_i^{-} \to c^{\dag}_i$, $\sigma_i^{+}
\to c_i$, and $\sigma_i^{z} \to 1-2 c_i^\dag c_i$. By applying the
Fourier transform to the chain operators, we obtain

\begin{eqnarray}
H&{=}&\sum_{k}\epsilon _{k}c_{k}^{\dagger }c_{k}
-h_S(1-2c^\dag_S c_S) -  \sum_{i=1}^n h_{R_i}(1-2c^\dag_{R_i} c_{R_i})\nonumber \\
&-&\frac{g}{\sqrt{N}}\sum_{k}[ c_{k}^{\dagger }(  e^{ik l_{S}}
c_{S}+ \sum_{i=1}^n e^{ik l_{R_i}}c_{R_i}) +H.c.],
\end{eqnarray}
where $k=2 \pi q/(N a)$, $a$ being the lattice constant and $q$ an integer number,  $\epsilon _{k}{=}-2h- 4 J \cos( k a)$ and
\begin{equation}
c_{k}{=}\frac{1}{\sqrt{N}}\sum_{l{=}1}^N c_l e^{i k l}.
\end{equation}
Without loss of generality, we will assume $a$ and $4J$,
respectively, as the units of length and energy ($\hbar$ is, as
usual, the unit of action).

The goal of a QST protocol is to act over an initial state encoded
in the spin at the sender site with both the set of the receivers
and the channel aligned in  state
 $|\phi_{in}\rangle{=}(\alpha |0\rangle+\beta|1\rangle)_S   |0\rangle_R^{\otimes n} |0\rangle_C^{\otimes N} $,
and, by exploiting the dynamical evolution for a definite transfer
time, transform it into $|\phi_{out}\rangle{=} |0\rangle_S (\alpha
|0\rangle+\beta|1\rangle)_{R_j}   |0\rangle_{ \bar{R}_j }^{\otimes
n-1}  |0\rangle_C^{\otimes N} $,  where $\bar{R}_j$ is the
register of the $n-1$ receivers complementary to $j$.
Since
$|0\rangle_S |0\rangle_R^{\otimes n}  |0\rangle_C^{\otimes N}$
is an eigenstate of $H$, it will be enough for our purpose to study the conditions under
which  $|1\rangle_S |0\rangle_R^{\otimes n}|0\rangle_C^{\otimes N} $ evolves into
$|0\rangle_S |1\rangle_{R_j}|0\rangle_{ \bar{R}_j}^{\otimes n-1}|0\rangle_C^{\otimes N} $
or, in the language of fermion excitation introduced before, we
want to know if there exists a time $t^*$ such that {$
c_S^\dagger(t^*)|0 \rangle{\simeq}c^\dagger_{R_j} |0\rangle$.
In order to get a full characterization of the QST, one should
evaluate a fidelity averaged over all of the possible initial
states (that is, over all the possible combinations of $\alpha$
and $\beta$ such that $|\alpha|^2+|\beta|^2{=}1$). It has been
shown in Ref. \cite{bose2003} that this average fidelity only
depends on the transition amplitude $f_{R_jS}$ of an excitation
from the sender to the $j$-th receiver, through the relation
$\bar{F}{=}\frac{1}{2}+\frac{|f_{R_jS}|}{3}+\frac{|f_{R_jS}|^2}{6}$.
As a result, for a generic $R_j$, the average fidelity is a
monotonous function of the transition probability
$F_{R_j}(t){=}|f_{R_jS}|^2 \equiv |\langle 0|c_{R_j}
c^{\dagger}_s(t) |0\rangle|^2$, and $\bar F$ reaches unity only
for $F_{R_j}(t){=}1$. We can, therefore, consider the behavior of
$F_{R_j}$ itself and, as we want to route the information to many
receivers, the protocol should be able to guarantee the highest
possible value for this probability, independently of the selected
receiver's location.

The dynamical problem is completely specified by the following set of coupled equations for the $N+n+1$ variables:
\begin{eqnarray}
\dot{c}_S^\dag&{=}&-i\Omega_S c_S^\dag      +i\frac{g}{\sqrt{N}}\sum_k c_k^\dag, \label{eqc}\\
\dot{c}_{ R_{i}}^\dag&{=}&  -i\Omega_{R_{i}} c_{R_{i}}^\dag      +i\frac{g}{\sqrt{N}}\sum_k e^{-ik l_{R_{i}}} c_k^\dag, \label{eqr}\\
\dot{c}_k^\dag&{=}&-i \epsilon_k c_k^\dag +i\frac{g}{\sqrt{N}}\left(c_S^\dag+\sum_{i=1}^n  c_{R_{i}}^\dag e^{ik l_{R_{i}}}\right), \label{eqs}
\end{eqnarray}
where we have used the notation $c_j^\dag{=}c_j^\dag(t{=}0)$,
$\Omega_j=-2h_j$ ($j{=}k,S,R_i$) and have assumed that $l_S{=}0$. As
discussed in Ref.~\cite{bus}, in the weak coupling limit, a
solution can be worked out in the Laplace space and then can be brought
back to the time domain.

In the following, we will describe how to obtain an efficient
routing within the model described so far.
\subsection{Chain-receivers resonance}\label{sec:res_weak1}
 In the scheme we are
proposing, we exploit the resonance between the local energy of at
the receiver site and one of the modes of the chain in order to
achieve the transfer. An efficient routing protocol, then,
requires that we are able to resolve the different levels of the
energy spectrum. To this aim, we must consider a finite size
system with a number of sites $N$ limited by the minimal relevant
energy separation that one is able to resolve.

To better illustrate our idea, we start by considering the ideal
case of a channel where all the energy levels are well separated
and resolved. The sender and the receivers are coupled to
different sites of the channel by a hopping term whose strength we
assume weak with respect to the intra-channel one. The local
energy of every receiver can be made resonant with a different
mode of the channel. In our specific case, since the channel
levels are twofold degenerate, with the exception of the $k{=}0$
and $k{=}\pi/2$ modes, the number of receivers can be, at most,
$N/2+1$. In the presence of a small hopping constant between
receiver $R_j$ and chain site $j$, the degeneracy is resolved, and
the resonant states are  split into two new levels separated by an
energy amount $\delta$. The weak-coupling condition holds when the
splitting is smaller than the original energy separation in the
chain.

Roughly speaking, the dispersion $\epsilon_{k}$ can be divided
into a parabolic region at the bottom and at the top of the energy
band and a linear region in the middle of the band. In the
parabolic region, the energy separation is on the order of
$\Delta_p \simeq \pi^2 /(2 N^2)$ whereas, in the linear region of
the band, $\Delta_l \simeq 2\pi /N$. Therefore,  weak coupling
conditions are fulfilled whenever $\delta \ll \Delta_p$ (see Fig. \ref{fsch1}). In this
way, every receiver is coupled to the channel only via its
resonant mode, whereas transitions via the other modes can be
neglected. The energies of the channel and the receivers are
fixed, whereas the sender can tune its energy. The sender selects
the receiver $R_j$ to send the state to, by tuning its energy
$\Omega_S$ to $\Omega_{R_{j}}$. In this way, the system behaves as
an effective model in which only the sender, the receiver, and the
resonant modes of the channel are involved in the dynamics. As
pointed out in Ref.~\cite{bus},  for a channel with an odd number
of sites, destructive interference occurs, and the excitation only
oscillates between sender and channel, without arriving at the
receiver. So, we will restrict ourselves to consider the case of
an even number of sites $N$. Moreover,   in order to achieve
efficient state transfer, the receiver has to be coupled  to  a
site with an even position label.
\begin{figure}
\begin{center}
  \includegraphics[height=4cm,angle=0]{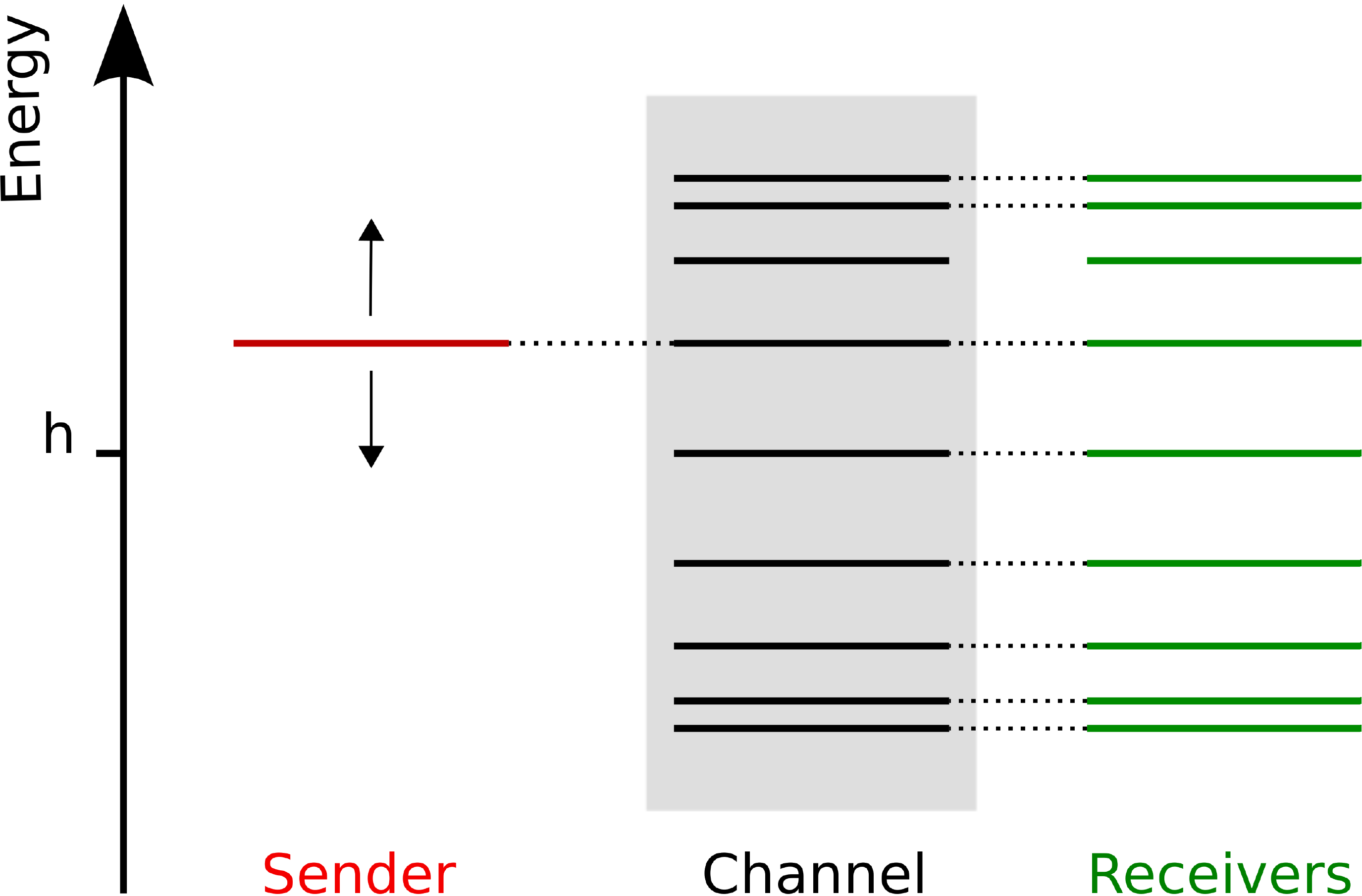}
\caption{(Color online) Energy-level scheme. Each of the receivers
is in resonance with one of the (pairs of) levels of the discrete
band. By locally tuning the sender energy, it is possible to
select the desired receiver. The spectral separation determines a
bound for the maximum number of receivers.}
 \end{center}\label{fsch1}
\end{figure}

Following the calculation of Ref.~\cite{bus}, a weak-coupling
expansion in $g$ can be performed to solve the system of
Eqs.~(\ref{eqc,eqr,eqs}). When
$\Omega{=}\Omega_S{=}\Omega_{R_j}$ is chosen to be resonant with
two modes $\pm \bar{k}$ of the channel, because of the
interaction, these four degenerate levels are split into
$\Omega\pm\delta_{\pm}$, where
\begin{equation}
\delta_{\pm}  \simeq   \frac{\bar{\omega}}{\sqrt{2}} \sqrt{1\pm\cos \bar{k} R_j},
\end{equation}
and where $\bar{\omega}{=}2g/\sqrt{N}$. The transition probability
for the receiver $R_j$, then,  has the form
\begin{equation}\label{eqn:fid}
F_{R_j}(t)\simeq \frac{1}{4}\left( \cos \delta_+ t -\cos  \delta_-t \right)^2.
\end{equation}
\begin{figure}
\begin{center}
  \includegraphics[height=7cm,angle=-90]{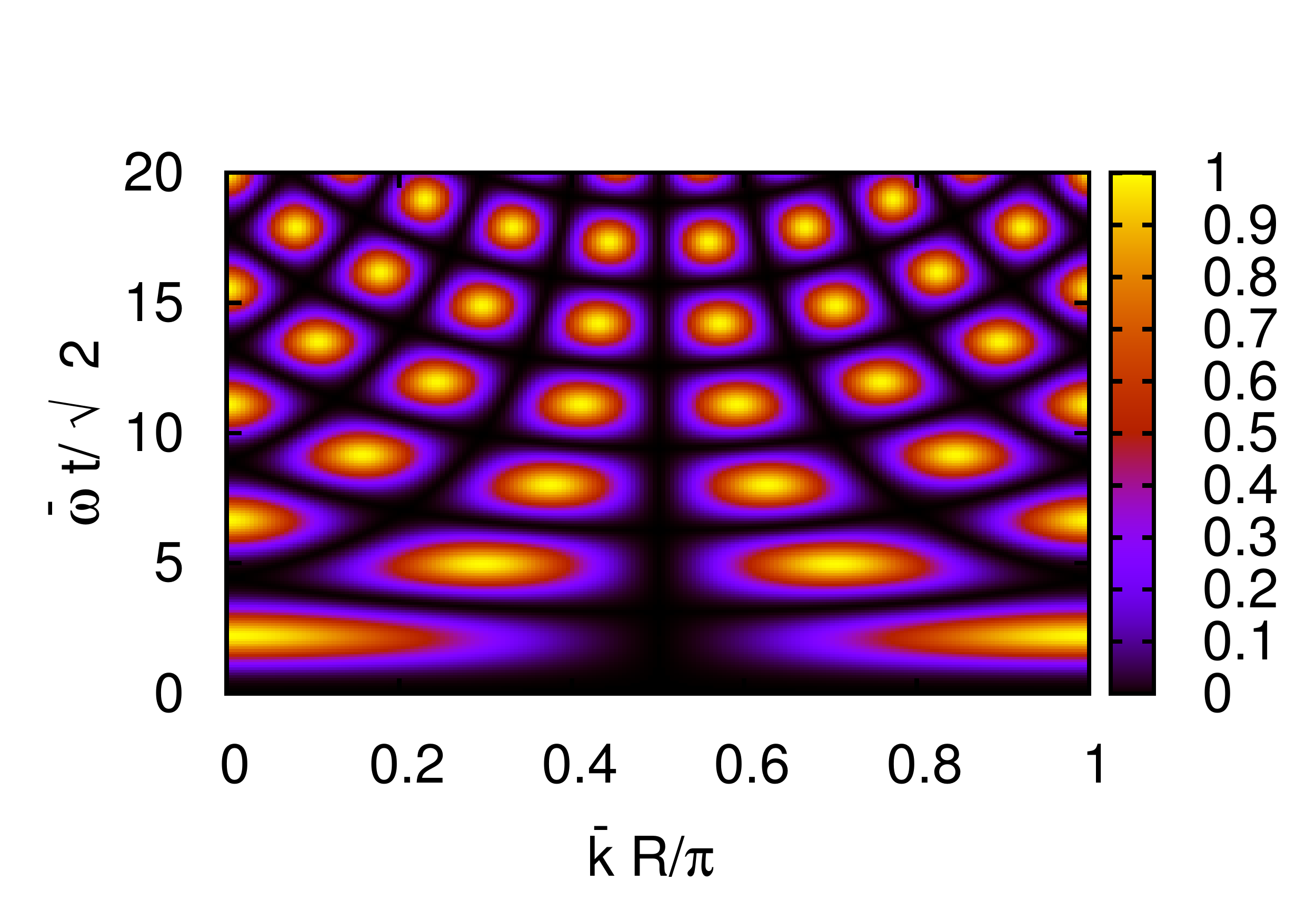}
\caption{(Color online) Transition probability as a function of
time (scaled with $\bar{\omega}$) and $\bar{k} R$. Integer values
of $\bar{k} R/\pi$ guarantee high quality transfer after shorter
times [yellow (light) spots in the density plot].}\label{ffid}
\end{center}
\end{figure}
As mentioned before, high-fidelity QST is achieved if, for a
certain time, $F_{R_j}$ approaches $1$. Since it depends on the
product $\bar{k} R_j$, then, it is clear that, for every position
around the chain, there is an optimal energy in the band spectrum.
In Fig.~\ref{ffid}, the transition probability (\ref{eqn:fid}) is
plotted. It appears evident that the choice $\bar{k} R_j{=}\pi s$,
with $s$ as an integer,  is always optimal since the time  $F_{R_j}$
takes to reach its maximum is shorter.

As shown in Fig.~\ref{ffid}, the maximum  can also be achieved  for
different choices of $\bar{k} R_j$ but only after longer times.
This means that decoherence effects, caused by the unavoidable
presence of some external environment, are more likely to come out.
As a consequence of  this environmental intrusion,
the quality of the routing protocol  can be seriously affected.

 For
$\bar{k}{=}\pm \pi/2$, which corresponds to the linear part of the
dispersion, Eq.~(\ref{eqn:fid}) reduces to
\begin{equation}\label{eqn:fid_lin}
F_{\rm{lin}}(t)\simeq \sin^4 \left( \frac{gt}{\sqrt{N}}\right).
\end{equation}
This case corresponds to the most efficient configuration, since
the energy separation with the closest levels is the highest.
Moreover, the value of $F$ is independent both of the
receiver's position and of the time for reaching the maximum.
Another important case is given by the resonance with the modes
$\bar{k}{=}0$ or $\bar{k}{=}\pi$ where the dispersion is
quadratic. This case cannot be deduced as a limit of
Eq.~(\ref{eqn:fid}) because these two modes are not degenerate.
Following the procedure of Ref.~\cite{bus}, one obtains
\begin{equation}\label{eqn:fid_quad}
F_{\rm{quad}}(t)\simeq \sin^4 \left(\frac{g  t}{\sqrt{ 2 N}}\right).
\end{equation}
In this case too, $F$  does not depend on $R_j$, but here, the
energy separation is smaller, and other levels could couple to the
dynamics, making the perturbative approximation less accurate.

In Fig.~\ref{fcfr}, the exact numerical evaluation of the
transition probability (calculated for a chain of $N=16$ sites,
assuming $g{=}10^{-2}$) is compared with the analytical
expressions given in
Eqs.~(\ref{eqn:fid,eqn:fid_lin,eqn:fid_quad}).
\begin{figure}
\begin{center}
  (a)
  \includegraphics[height=7cm,angle=-90]{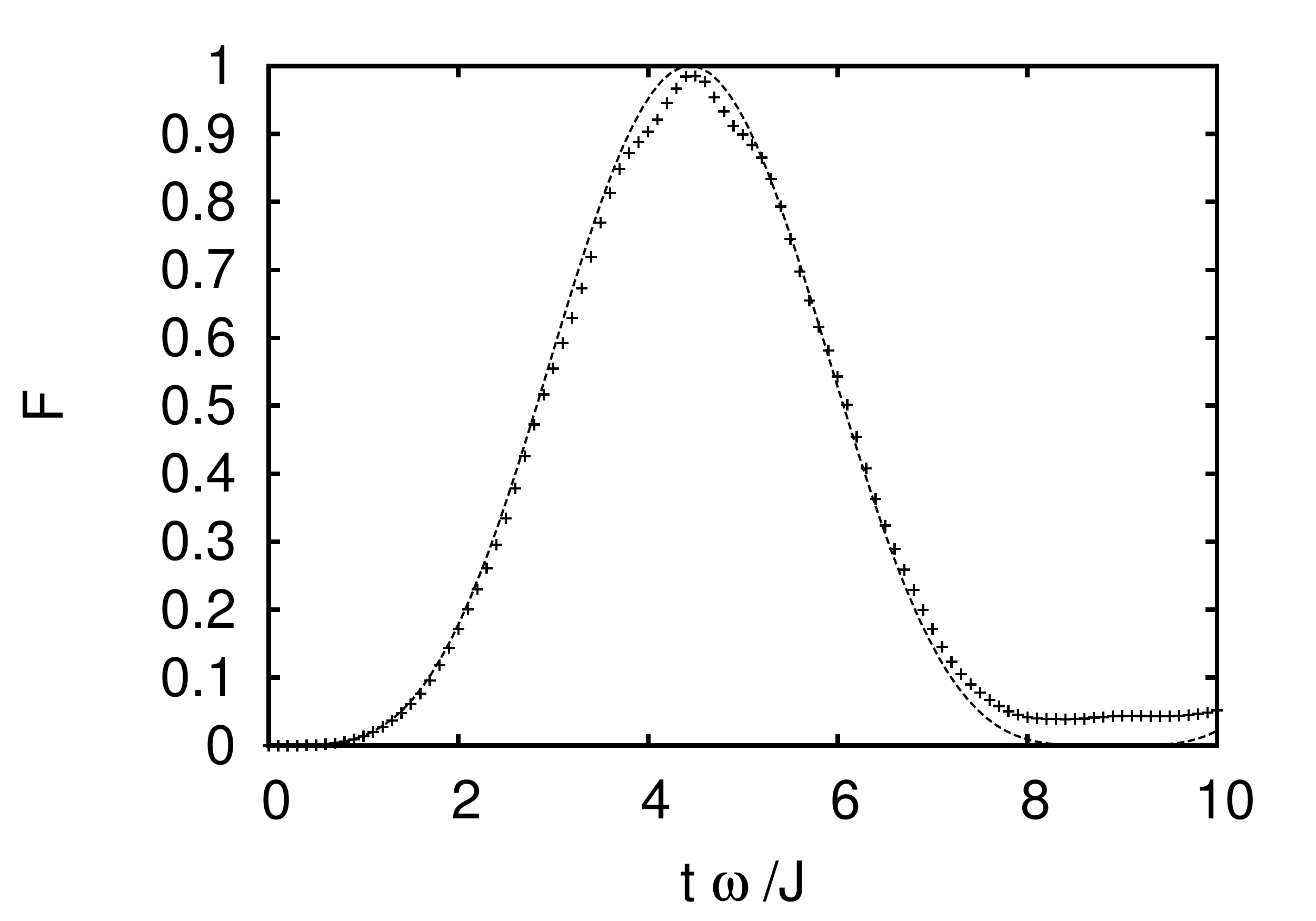}\\
  (b)
  \includegraphics[height=7cm,angle=-90]{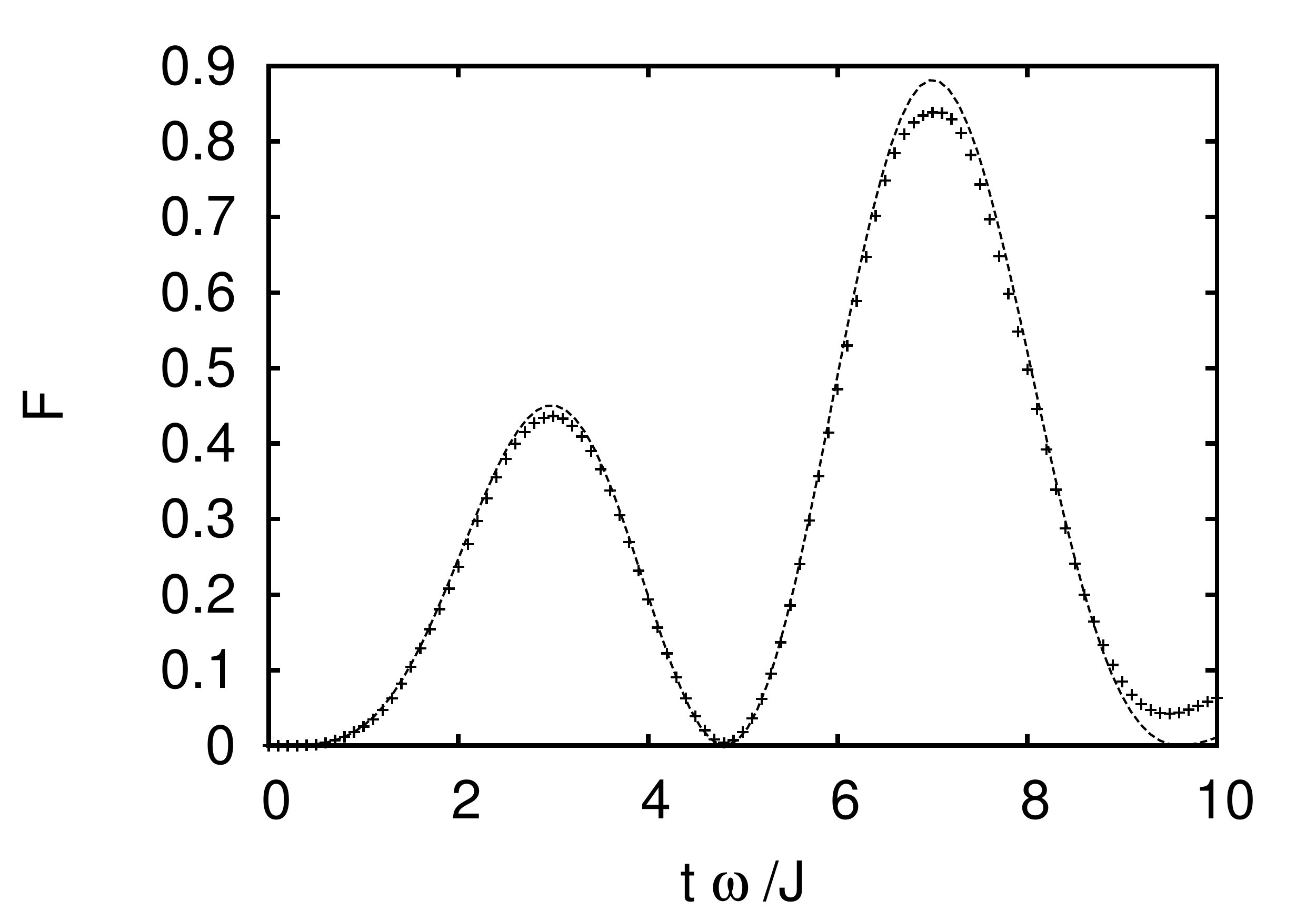}\\
  (c)
  \includegraphics[height=7cm,angle=-90]{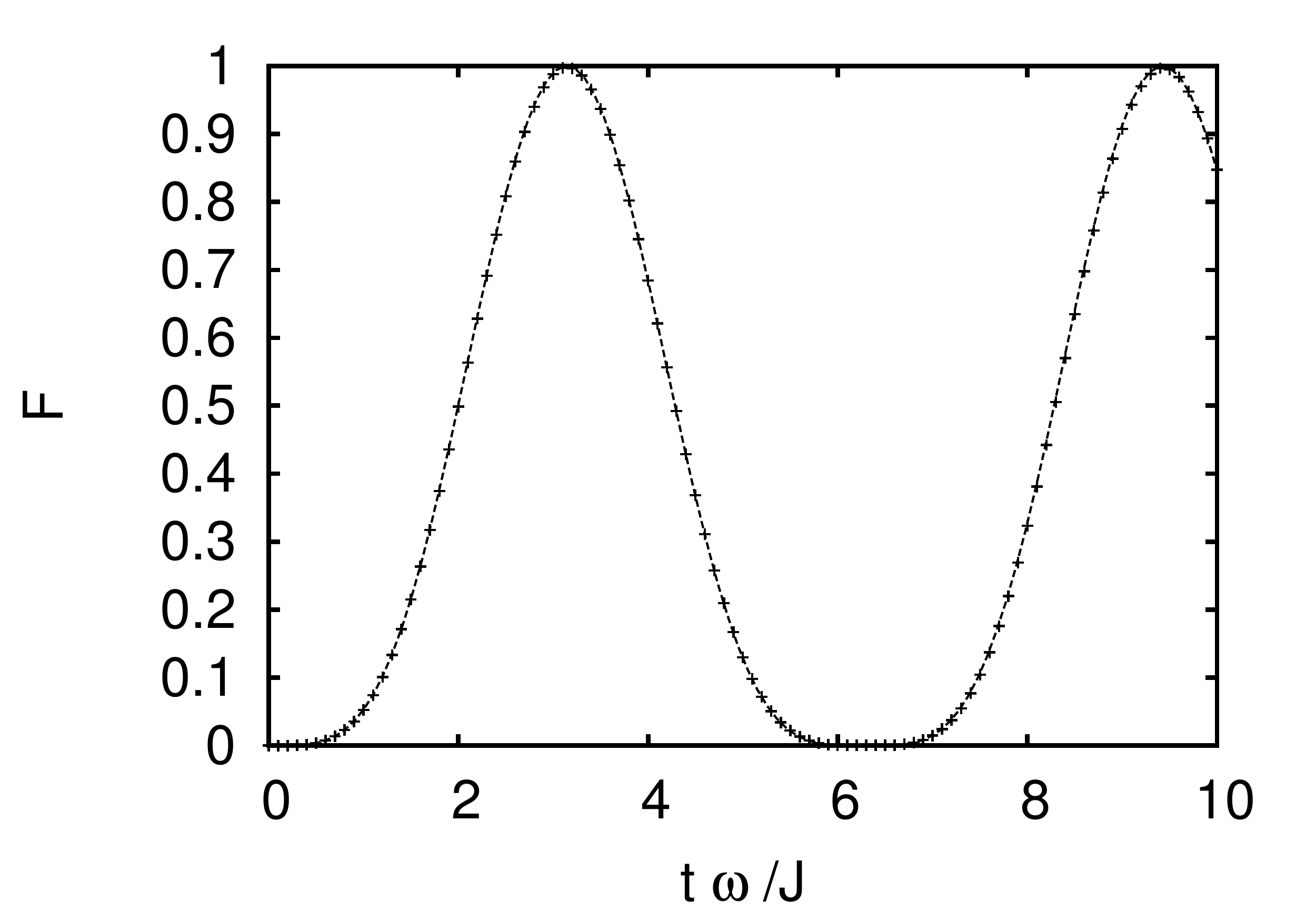}\\
\end{center}
\caption{(Color online) Comparison of the exact transition
probability $F$ (dashed line) with the analytic result obtained by
applying the weak-coupling approximation (continuous line) for
$N{=}16$ and $g{=}0.01$. (a) $\bar{k}{=}0$ and $R{=}12$;    (b)
$\bar{k}{=}7 \pi/4$ and $R{=}10$; (c) $\bar{k}{=} \pi/2$ and
$R{=}4$. }\label{fcfr}
\end{figure}
The best efficiency is achieved for those receivers that are
resonant with the linear and quadratic parts of the dispersion,
while $F$ is reduced in the intermediate cases.

The need to be resonant with a mode which lies in one region or
another of the spectrum of the chain, by itself, introduces an
inhomogeneity among the receivers. Moreover, the energy separation
in the quadratic part of the dispersion decreases very rapidly
with the number of sites, posing a limit to the length of the
channel (see Fig. \ref{scheme2}). In principle, this problem could be overcome by
decreasing the coupling $g$, but this would imply longer
transmission times, and  the decoherence effect would start to be
relevant.

\begin{figure}
\begin{center}
  \includegraphics[height=4cm,angle=0]{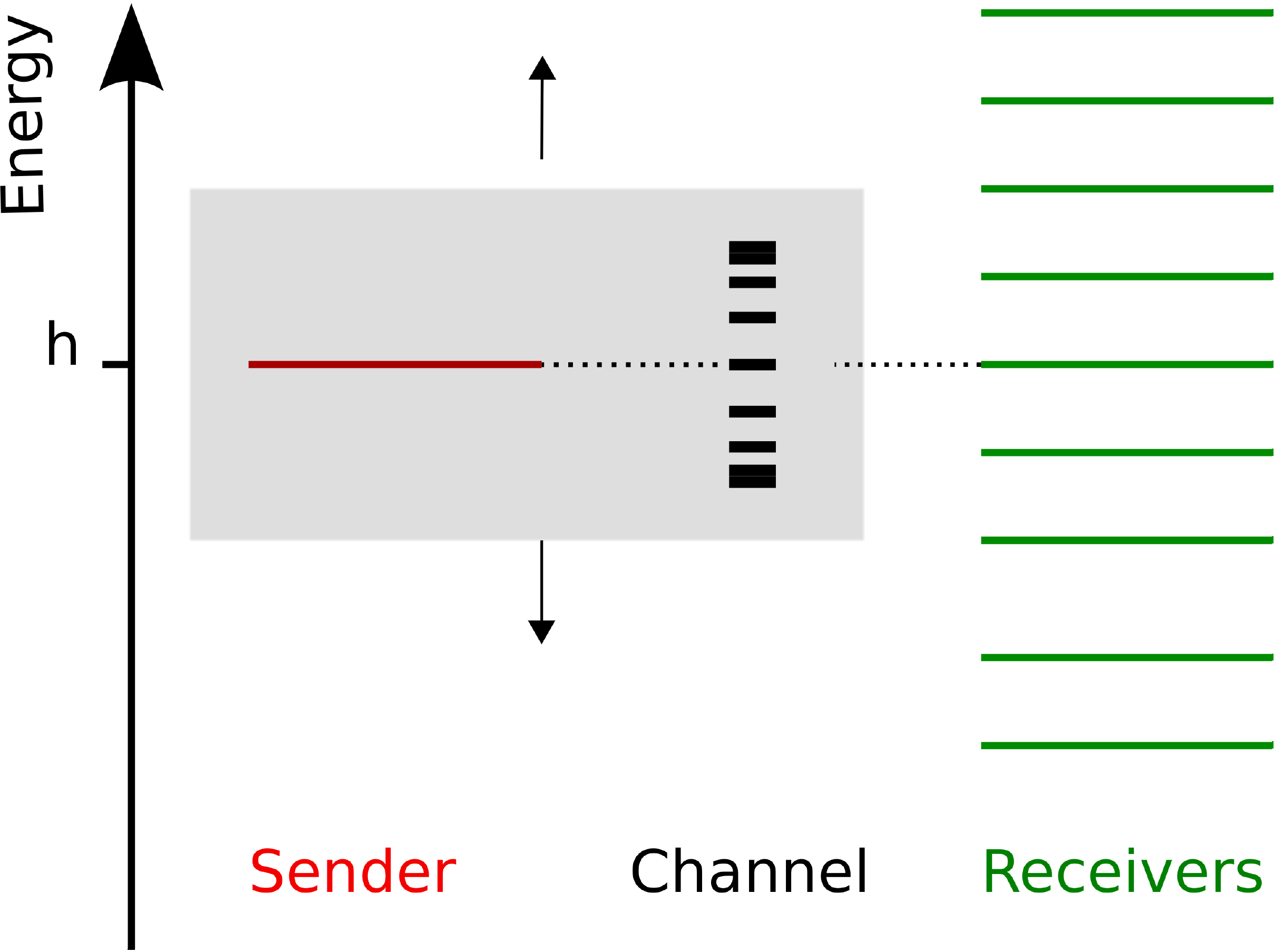}
\caption{(Color online) Modified scheme of levels. The receivers
are well separated in energy ($\Delta_R >2\pi /N $) by tuning the
sender and by translating the whole chain spectrum through a
uniform, external magnetic field.}\label{fsch2}
\end{center}
\end{figure}
\subsection{ Equally spaced energies}\label{sec:res_weak2}
In order to increase the
possible number of receivers and to get an equivalent fidelity for
each of them, a different energy configuration can be considered.
As sketched in Fig.~\ref{fsch3}, let us assume the sender to always be
 resonant with the mode $\bar{k}{=}\pi/2$ in the linear
dispersion region, so that the dispersion of the channel is
$\Omega_S-\cos \bar{k}$. Let us also assume that the energies of
the receivers do not match the band levels but are separated by
$\Delta_R >2\pi /N $.

A receiver $R_j$ is selected by tuning $\Omega_S{=}\Omega_{R_j}$
(and by changing, accordingly, the field on the chain). Because of
the validity of the weak-coupling approximation, all of the other
receivers are not involved in the process. Given that we are
working in the linear dispersion region, the transition
probability is given by Eq.~(\ref{eqn:fid_lin}).

This improves the previous scheme since the effects of the
quadratic part of the  band are now corrected, and the fidelity is
the same for every receiver.

Since the typical energy separation is now the one in the middle
of the band,  for a fixed value of $g$, longer chains can be
employed, and a larger number of receivers can be included. However,
for this scheme to work, it is not sufficient to act only on the
sender any more, but a global control over the chain is necessary.
This could be obtained by applying a global magnetic field that
has the effect of translating the whole band spectrum by the
desired amount. As in the former proposal, no control over the
receivers is needed.
\begin{figure}[h!]
\begin{center}
  \includegraphics[height=4cm,angle=0]{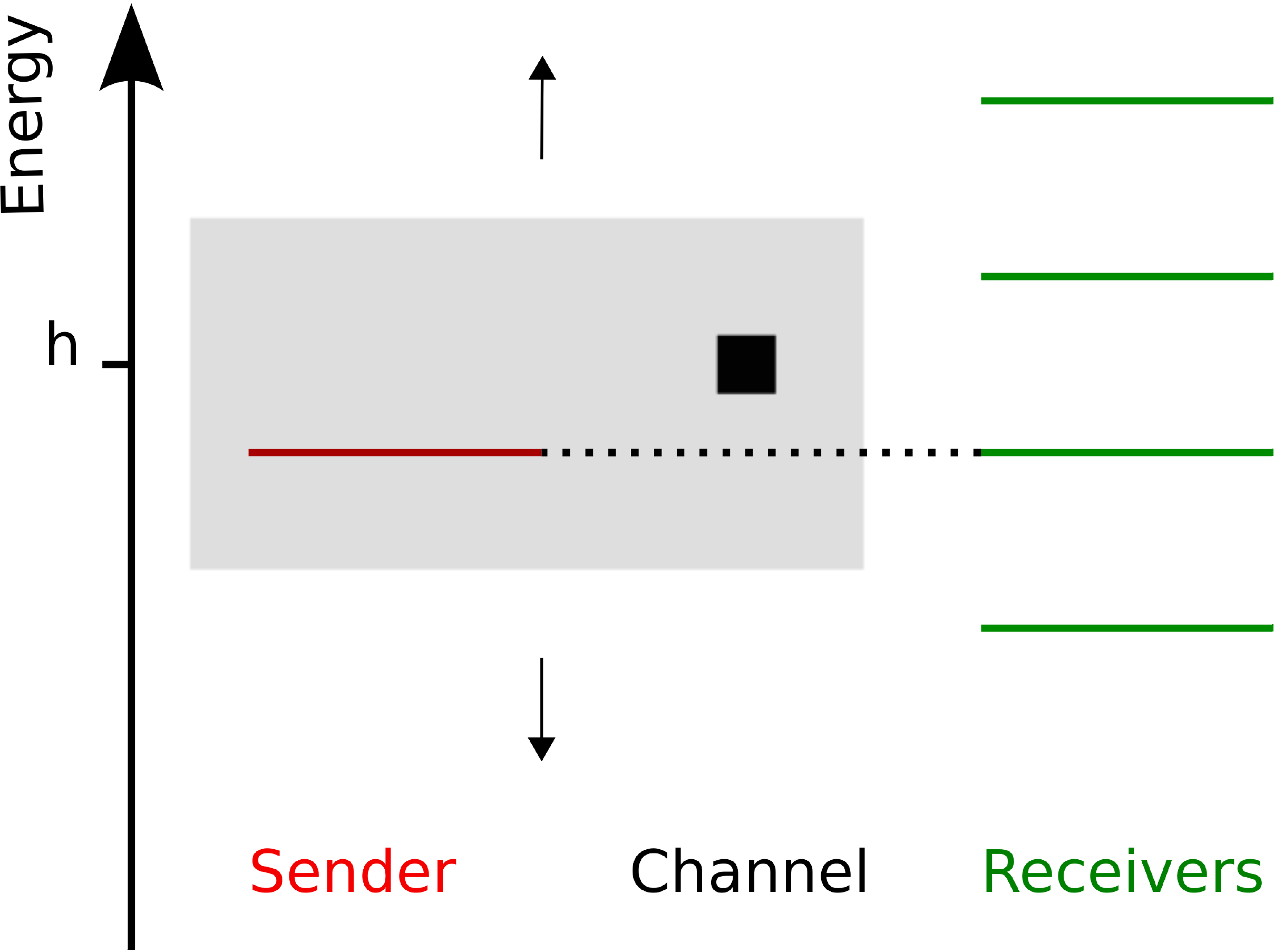}
\caption{(Color online) The sender and one of the receivers are in
resonance, whereas the continuous (as seen from outside) channel is
detuned. The effective two-body frequency oscillation associated
with this scheme is lower than the other schemes proposed in this
section.}\label{fsch3}
\end{center}
\end{figure}

\subsection{ Off resonance}\label{sec:res_weak3}
Finally, we just mention a third
possible scheme consisting of an almost continuous channel with a
weak out-of-resonant coupling with the sender and the
receivers~\cite{neto,bus}. In this scheme, the off resonant
continuous channel  creates an effective coupling between sender
and receiver tuned at the same energy. The setup is similar to the
previous one, but now $ h=\Omega_S \pm \nu$ with $\nu > 1$. This
condition ensures that the sender is not resonant with the
channel. As  for the receivers, the energy separation condition
becomes $\Delta_R > 2 $. This scheme allows for a high-fidelity
 transfer over longer distances, since, at least within the limits of validity of  the weak
coupling approximation, the system undergoes an effective two-level oscillation
between sender and receiver, whilst the chain, which is never populated, acts as 
a mere connector.
 As a drawback, longer times are required to accomplish the protocol
and, as in the chain-receiver resonance case, environmental
decoherence effects are more likely to affect the quality of the protocol.

\section{Quantum router via local field barrier}\label{sec:2model}
In the model presented in Sec.~\ref{sec:model}, we were supposed to be
able to reduce the coupling between the chain and the
sender-receiver sites. In this section we propose an alternative
configuration where the hopping is assumed to be equal between all
the spins, and routing is performed by tuning the local magnetic
field acting on the spin adjacent to the sender.
As there is no need to operate on the sender and/or the receiver
couplings, this may result in a simpler implementation depending
on the experimental set up.

Let us consider a linear $XX$ chain composed by $N$ sites plus $n{+}1$ pairs of spins.
Following the notation in Sec.~\ref{sec:model}, we can write the total Hamiltonian
as
$H{=}H_C{+}\sum_X H_{I_X}{+}\sum_X H_{{CI}_{X}}$, where
\begin{align}
&H_{I_X}{=}-J(\sigma^x_{X_A} \sigma^x_{X_B}+\sigma^y_{X_A} \sigma^y_{X_B})-h_{X}\sigma^z_{X_B},\notag\\
&H_{{CI}_{X}}{=}-J(\sigma^x_{l_{X}} \sigma^x_{X_B}+\sigma^y_{l_{X}} \sigma^y_{X_B}),
\label{H_lin}\end{align}
and open boundary condition of $H_C$ are assumed.
Here $X_{[A,B]}$ stands for the spins composing the sender block $S_{[A,B]}$ and the receiver blocks $R_{k[A,B]}$.

Each block is composed by a pair of spins: the first one, labelled
by $A$, acts as the effective sender/receiver and the second one,
labelled by $B$, is connected with the site $l_{X}$ belonging to
the linear chain.
\begin{figure}[h!]
\includegraphics[width=8.5cm]{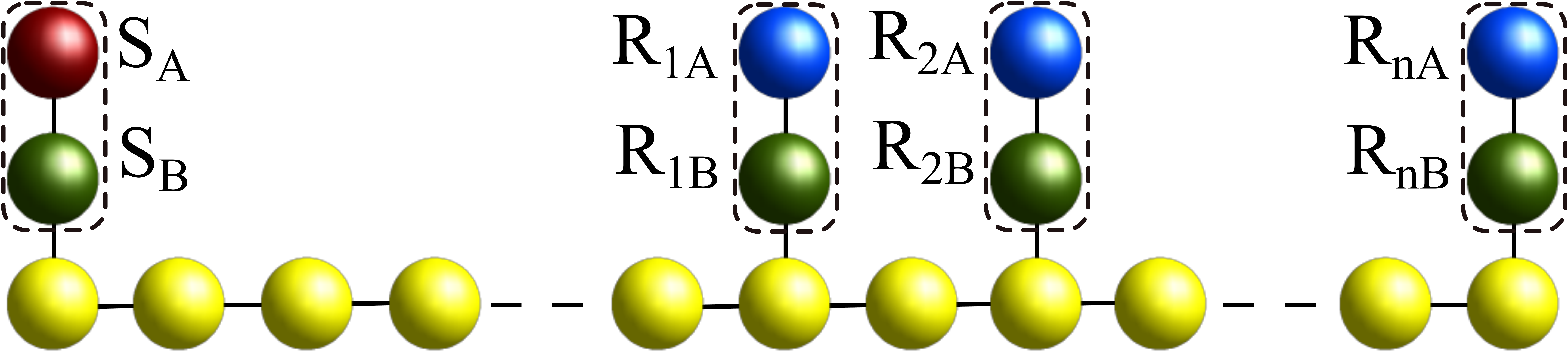}
\caption{(Color online) The model:  A linear $XX$ chain is used as a
transmission channel between the sender $S_A$ and one of the receivers $R_{kA}$.}
\label{fline}
\end{figure}

We assume that magnetic fields with (different) intensities
$h_{X}$ act on the second site of each block. By controlling
$h_{S}$, it is possible to confine the excitation on the sender or
to perform QST from $S_A$ to $R_{kA}$ by choosing
$h_{S}{=}h_{R_{k}}$. As specified in Ref. ~\cite{catena2_N-1}, for even
chains the optimal transfer time $t^*$ is proportional to the
square of the intensity of the magnetic field $h_{S}$ and, for
large enough $h_{S}$, it is also independent of the number of
sites. Therefore we will consider a configuration of the router in
such a way that there are an even number of spins between the
sender and each receiver, as depicted, e.g., in Fig.~\ref{fline}.

As in Sec.~\ref{sec:model}, we assume the initial state to be
prepared with all spins in the down state,
$\ket{\mathbf{0}}{=}\ket{0}^{\otimes N+2(n+1)}$. Then, we prepare
the sender site $S_A$ in the state
$\ket{\psi_{in}}{=}\alpha\ket{0}+\beta\ket{1}$ and let the
complete system evolve according to Eq.~\eqref{H_lin}. Because of
the invariance of the subspace with a fixed number of flipped
spins, the  fidelity averaged over all possible initial states
 is again given by the expression
$\bar{F}{=}\frac{1}{2}+\frac{|f_{RS}|}{3}+\frac{|f_{RS}|^2}{6}$ in Ref.~\cite{bose2003}, and the transition amplitude $f_{RS}$ reads
\begin{equation}\label{fRS}
f_{R_{jA},S_A}(t){=} 
\sum_{k{=}1}^{N{+}2(n{+}1)}\langle\mathbf{R}_{jA}|\mathbf{a}_k\rangle  \langle\mathbf{a}_k|\mathbf{S}_A\rangle e^{-i \lambda_k t}
\end{equation}
and $\lambda_k$,
$|\mathbf{a}_k\rangle{=}\sum_{j{=}1}^{N{+}2(n{+}1)}a_{kj}|\mathbf{j}\rangle\,
, \label{kbasis}$ are, respectively, the eigenvalues and the
corresponding eigenvectors of $H$ written in the position basis
$\ket{\mathbf{j}}{=}\ket{0\ldots 01_j 0\ldots 0}$ (with
${\mathbf{j}}{=}1,\ldots,N,S_{[A,B]},R_{1[A,B]},...,R_{n[A,B]}$),
where the spin at the $j$th site has been flipped to the $\ket{1}$
state. In order to perform an efficient QST in the setting under
scrutiny, it is necessary to achieve a modulus of the transmission amplitude
between sites $S_A$ and $R_{jA}$ as close as
possible to $1$ at a certain time $t^{*}$. The local field $h_X$
exactly produces  this result.

Indeed, the presence of $h_{X}$ has two consequences: First, it
causes the appearance of an eigenstate localized on the sites $B$
of each block, with energy much larger than that of the rest of
the system; and, second, an effective weak coupling of the spin at
sites $A$ of each block to that at site $l_{X}$ of the linear
chain arises.

This can easily  be seen by writing $H_{{CI}_X}$ as
\begin{equation*}
H_{{CI}_X}{=}-\frac{2J}{\omega^a_X-\omega^{b}_X}\left(\omega^a_X\ket{\mathbf{l}_X}
\bra{\psi^a_X}{-}\omega^b_X\ket{\mathbf{l}_X}\bra{\psi^b_X}+h.c.\right)
\end{equation*}
where $\omega^{a,b}_X$ and $|\psi^{a,b}_X\rangle$ are the eigenvalues and eigenvectors of $H_{{I}_X}$, after rescaling the ground state energy:
\begin{align*}
\omega^a_X{=}{-}h_X{+}\sqrt{h_X^2+4J^2}\; ; \;\ket{\psi^a_X}{=}\frac{\omega_X^b}{2J}\ket{\mathbf{X}_A}{+}\ket{\mathbf{X}_B},\\
\omega^b_X{=}{-}h_X{-}\sqrt{h_X^2+4J^2}\; ; \;\ket{\psi^b_X}{=}\frac{\omega_X^a}{2J}\ket{\mathbf{X}_A}{+}\ket{\mathbf{X}_B}.
\end{align*}
In the limit $h_X{>>}J$, the eigenstates $\ket{\psi^a_X}$ and $\ket{\psi^b_X}$
become $\ket{\mathbf{X}_A}$ and $\ket{\mathbf{X}_B}$,
the scaling of their coupling to the chain's site behaves as
\begin{equation*}
\frac{\omega^a_X}{\omega^a_X-\omega^b_X}\rightarrow\frac{J^2}{h_X^2}
\qquad\text{and}\qquad
\frac{\omega^b_X}{\omega^a_X-\omega^b_X}\rightarrow-(1-\frac{J^2}{h_X^2}).
\end{equation*}
It follows that we can write
\begin{equation*}
H_{{CI}_X}{=}-2J\left[\frac{J^2}{h_X^2} \ket{\mathbf{l}_X}\bra{\mathbf{X}_A}+(1{-}\frac{J^2}{h_X^2})\ket{\mathbf{l}_X}\bra{\mathbf{X}_B}+ h.c.\right].
\end{equation*}
\begin{figure}[h!]
\includegraphics[width=7cm]{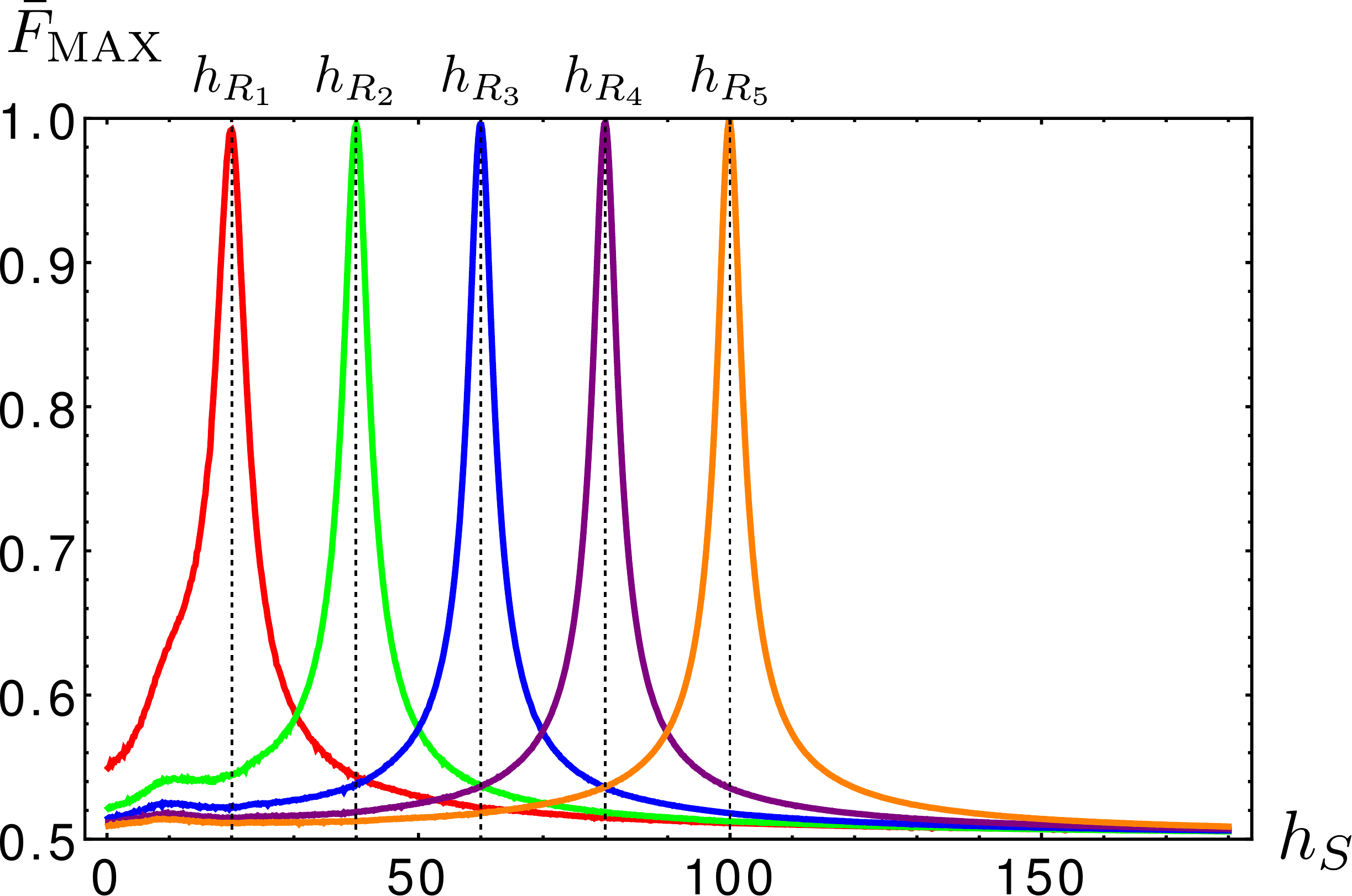}
\caption{(Color online) Maximum of the average fidelity in a fixed
time interval $Jt < 5 \times 10^4$, for a channel of $N=30$ sites
with $n=5$ receiver blocks. By tuning the magnetic field $h_{S}$
respectively to $h_{R_k}$, $k{=}1,...,5$, we can perform a routing
with high efficiency. The magnetic fields are ordered according to
the position of the receivers,
$h_{R_{(k-1)}}{<}h_{R_{k}}$.}\label{fresonance}
\end{figure}
This implies that we can effectively consider the first spin of
each block weakly coupled to the chain's spin with strength
${\sim}1/h^2_X$; whereas, the second spin, still coupled with
strength ${\sim}J$ to the chain, experiences the large magnetic
field $h_X$ of $H_{I_X}$, which freezes its dynamics.

As in the case of Sec.~\ref{sec:res_weak3}, when $h_{S}$ is close
to the energy of receiver block $h_{R_{k}}$, there exists a pair
of eigenenergies outside the spectrum of the chain whose
corresponding eigenstates are localized (symmetrically and
anti-symmetrically) on the $B$-parts of each block involved in the
transfer \cite{bus,local}. Moreover, in this case we have the
emergence of another quasi-degenerate pair of eigenvalues (inside
the energy band of $H_C$, but out-of-resonance with any of its
eigenvalues), whose corresponding eigenvectors have a
non-negligible superposition with the states $\ket{\mathbf{S}_A}$
and $\ket{\mathbf{R}_{kA}}$, so that they give rise to an
effective Rabi-like oscillation mechanism of the spin-excitation
between the sender and the selected receiver site of the router.
As a result, both the transition amplitude and the average
fidelity become very close to unity at half the Rabi period.

In Fig.~\ref{fresonance} we report the maximum of the average
fidelity over all initial states within a fixed time interval $Jt
<5 \times 10^4$ for a channel of $N=30$ sites with $n=5$ receiver
blocks: It is clearly shown that by properly tuning the magnetic
field $h_{S}$, one can perform a QST with high efficiency towards
each of the targeted receiving sites.

\begin{figure}[htbp]
\includegraphics[width=\linewidth]{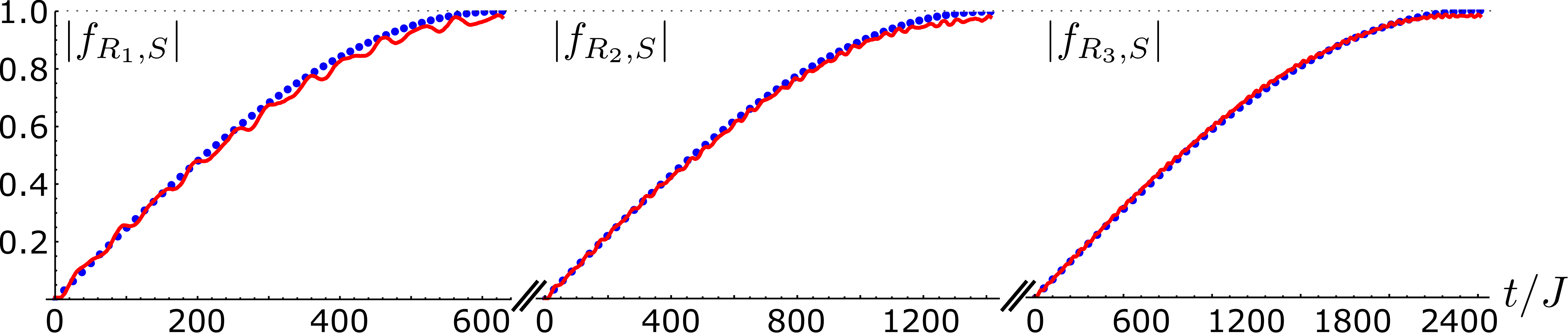}
\caption{(Color online) Comparison of the exact transition
amplitude in Eq.~(\ref{fRS}) (red solid line) with the result of
Eq.\ref{fapprox} (dotted blue line) obtained after a perturbative
analysis for $h_X\gg J$ in a chain of $N=20$ sites with $n=3$
receiver blocks. (Time is given in units of $1/J$)}
\label{ftrans_ampl}
\end{figure}
As shown in Ref.~\cite{catena2_N-1}, the transition amplitude
between the sender and a receiver, connected by an even number of
sites, is well approximated by
\begin{equation}
f_{R_{jA},S_A}(t)\approx\sin\left(\dfrac{J^3t}{h_j^2}\right), \label{fapprox}
\end{equation}
as checked in Fig.~\ref{ftrans_ampl} against the numerical
solution, for a chain of $N{=}20$ sites with $n=3$ possible
receiving blocks.

\section{Conclusions}\label{sec:concl}

The implementation of many quantum information protocols requires
the transfer of a quantum state from an input to different output
locations, and quantum routing has to be implemented in order to
build a large network. Depending
on the physical system used for this purpose, the control
over many interaction parameters may be unfeasible, and it
is necessary to study efficient routing protocols that require
minimal engineering and external manipulation.

In this paper, we have presented two different possible
implementations of a router that allows quantum state transfer
from a sender to a chosen receiver by means of a resonant
coupling mechanism. In the first scheme, the key ingredient
is the weak coupling between the sender and receivers to the
spin bus, and three different configurations of local and global
magnetic fields are considered: (a) Every receiver energy is
resonant with a different mode of the channel, and QST occurs
by tuning the sender energy; (b) the energies of the receivers do
not match the band levels but are equally spaced, and the sender
is always resonant with a mode of the channel so that QST can
be performed by translating the whole band spectrum by the
desired amount of energy through the application of a global
magnetic field; (c) a weak, out-of-resonance coupling between
the chain and the sender and the receivers. In this scheme, the
off-resonant continuous channel creates an effective coupling
between the sender and the receiver,
provided they are tuned to the same energy.

Finally, for the case in which the couplings between
adjacent qubits are constrained to be equal, we have proposed
a second model for the quantum routing protocol in which a
linear chain is used as a data bus, and the single sender and receiver spins are substituted by sender and receiver blocks
made of pairs of spins. One of these two spins is effectively
involved in the communication, whereas, the second (the
barrier spin, effectivelyworking as a gateway), is acted upon by
a local fieldwhich plays the role of a knob that permits theQST.
As a consequence of the use of strong local magnetic fields, an
effective weak coupling is established either between sender
and receiver and spin bus (in the resonant case) or between the
sender and the receiver (in the off-resonant case). Moreover,
the presence of the barrier spin makes it not necessary to act
directly on the sender qubit, which is, therefore, involved only
in the state encoding step, whichmay result in an experimental
simplification.

In the resonance regime, the information transfer is due
to collective degrees of freedom (i.e., the single-particle
excitations of the spin chain), and therefore, to obtain a good
transmission performance, it is necessary to be able to set the
energy levels in a precise way. This kind of control can be
achieved in the context of atomic Mott insulators where it
has been shown experimentally that different lattice potentials
can be tailored with high accuracy \cite{weitenberg}. On the other hand,
by working in the off-resonance regime, the precise shaping
of the mediumÕs energy level is unnecessary, and naturally
occurring systems as well as separated nitrogen vacancy
centers in diamond would represent a feasible experimental
implementation~\cite{AjoyC2013}.

Finally, since our theoretical treatment exploits a model
Hamiltonian, which received much experimental attention in
the past few years, it is definitely worthwhile to investigate
routing implementations based on it. These could be further
developed and improved in various ways; in particular, by
allowing for the possibility of multiple sending sites, connected
at will to a selected set of receivers, in order to perform multiple
quantum state transfer over a single data bus.

\acknowledgments
Financial support by the Science Foundation of Ireland under Project
No. 10/IN.1/I2979 is acknowledged. T.J.G.A. is supported by the
European Commission, the European Social Fund, and the Region
Calabria through the program POR Calabria FSE 2007-2013-Asse IV
Capitale Umano-Obiettivo Operativo M2. S.P. was supported by the
Spanish Ministry of Science and Innovation through the program
Juan de la Cierva. We acknowledge support from the Spanish MICINN
Grant No. FIS2008-01236, Generalitat de Catalunya (Grant No. SGR2009:00343),
Consolider Ingenio 2010 (Grant No. CDS2006-00019), European Regional
Development Fund, the Brazilian Ministry of Science and Technology (MCT) and the Conselho
Nacional de Desenvolvimento Cientifico e Tecnologico (CNPq).

\end{document}